\documentclass[aps,prx,reprint,twocolumn,superscriptaddress,nofootinbib, showkeys]{revtex4-2}
\pdfoutput=1
\usepackage[utf8]{inputenc}
\usepackage[english]{babel}
\usepackage{amsmath, amssymb}
\usepackage{graphicx} 
\usepackage{quantikz} 
\usepackage{float}
\usepackage{graphicx}
\usepackage{hyperref}
\usepackage{booktabs}
\usepackage{xcolor}

\usepackage{soul} 

\def\uimp{U_\text{imp}}

\begin{document}

\title{Physics-Informed Variational Quantum Classifier for Phase Detection \\ in Strongly Correlated Matter}

\author{Hugo Catalá}
\email{hugocatalacalatayud@gmail.com}
\affiliation{Instituto de Física Corpuscular, Universitat de València – Consejo Superior de
Investigaciones Científicas, Parc Científic, E-46980 Paterna, Valencia, Spain.}
\affiliation{Escuela de Ciencias, Ingeniería y Diseño, Universidad Europea de Valencia, Paseo de la Alameda 7, 46010, Valencia, Spain}

\author{Ezequiel Valero}
\affiliation{Escuela de Ciencias, Ingeniería y Diseño, Universidad Europea de Valencia, Paseo de la Alameda 7, 46010, Valencia, Spain}
\affiliation{Facultat de Física, Universitat de València, Carrer del Dr. Moliner, 50, 46100 Burjassot, Valencia, Spain}

\author{Germán Rodrigo}
\affiliation{Instituto de Física Corpuscular, Universitat de València – Consejo Superior de
Investigaciones Científicas, Parc Científic, E-46980 Paterna, Valencia, Spain.}

\date{June 12, 2026}

\begin{abstract}
The characterisation of quantum phases in strongly correlated systems is a crucial milestone for the deployment of quantum sensors. In this work, we present a Physics-Informed Variational Quantum Classifier (VQC) designed to detect the topological phase transition between the Fermi polaron quasiparticle and the molecular bound state. Unlike conventional Machine Learning approaches, our quantum architecture is constructed via the Trotterised time-evolution of an effective Hamiltonian, ensuring that the learnable parameters correspond to interpretable physical quantities. We show that the VQC efficiently discovers the optimal interferometric protocol, specifically the evolution time and effective bath interactions required to maximise the visibility of Ramsey fringes, thereby clearly distinguishing the Bose-Einstein Condensate (BEC) and Bardeen-Cooper-Schrieffer (BCS) regimes. Furthermore, we report the validation of this classifier on the QRed superconducting quantum processor (BSC-CNS). Despite the intrinsic hardware noise and decoherence, the VQC preserves the relative ordering of the topological phases. We demonstrate that the physics-informed architecture achieves a linear gate complexity $\mathcal{O}(N)$, bypassing the exponential memory wall of classical simulation and ensuring scalability to many-body regimes.
\end{abstract}

\keywords{Quantum Machine Learning, Fermi Polaron, BEC-BCS Crossover, Superconducting Qubits, Physics-Informed Neural Networks, Ramsey Interferometry, NISQ}

\maketitle

\section{Introduction}

The physics of ultracold Fermi gases offers a unique controlled environment for investigating strongly correlated quantum matter~\cite{bloch2008many}. An important aspect in this field is the interplay between individual impurities, described by Fermi polarons~\cite{chevy2006universal,schirotzek2009observation}, and collective pairing phenomena, such as the crossover between Bose-Einstein Condensation (BEC) and Bardeen-Cooper-Schrieffer (BCS) superfluidity ~\cite{giorgini2008theory,regal2004observation}.

While the conceptual framework of the BEC-BCS crossover is well established, theoretical predictions for these strongly correlated dynamics face severe computational limitations, most notably the exponential memory wall and the fermionic sign problem. At the same time, experimental detection remains challenging, as conventional methods~\cite{ketterle2008making} rely on destructive measurements that discard critical phase information and are highly sensitive to signal degradation.

To bridge the gap between classical computational limits and experimental difficulties, Quantum Machine Learning (QML) emerges as a powerful tool. Variational Quantum Classifiers (VQC) leverage the high-dimensional Hilbert space of quantum processors to identify complex patterns in data~\cite{havlicek2019supervised,schuld2019quantum}. However, standard VQC approaches frequently employ generic quantum circuit Ans\"atze with no physical connection to the problem at hand. 

In a previous work~\cite{Catala2026}, we established an effective Hamiltonian formalism that bridges the BEC and BCS regimes, demonstrating that the transition from a dressed quasiparticle to a molecular bound state is governed by a continuous evolution of the population imbalance and the interaction strength.

In this work, we present a Physics-Informed VQC designed to accurately characterise the polaron-molecule transition. Rather than treating the quantum processor as a passive simulator, our approach leverages the VQC to engineer the optimal measurement setup by processing the quantum state directly within the Hilbert space, specifically Ramsey interferometry, required to bypass noise and detect phase transitions efficiently.

Our approach differs from standard Machine Learning  in that the VQC architecture is explicitly dictated by the time-evolution of the effective Hamiltonian. Consequently, the training process does not merely optimise abstract weights; rather, the VQC learns the optimal physical parameters, specifically the Trotterisation time step ($\delta t$) and the effective bath-bath interaction ($U_\text{ff}$) required to maximise the interferometric contrast between topological phases.

We report the successful demonstration of this quantum classifier on the QRed superconducting quantum processor at the Barcelona Supercomputing Center (BSC-CNS). We show that by exploiting the quantum interference patterns (Ramsey fringes) characteristic of the polaron dynamics, the VQC achieves high accuracy even in the presence of significant hardware noise.

The article is organised as follows. In Section~\ref{sec:PIQNNA}, we detail the architecture of the Physics-Informed VQC, deriving the quantum circuit Ansatz directly from the time-evolution of the effective Hamiltonian. Section~\ref{sec:STPD} describes the supervised training protocol using synthetic data and analyses the resulting quantum phase diagram. In Section~\ref{sec:EVQP}, we present the experimental validation on the QRed superconducting processor, discussing the impact of hardware noise. Section~\ref{sec:QVCA} explains the primary advantages of QML over classical approaches for these types of strongly correlated systems. Section~\ref{sec:SA} supports our work with a comprehensive scalability analysis. Finally, Section~\ref{sec:C} summarises our findings and outlines future directions for quantum sensing.

\section{Physics-Informed Variational Quantum Classifier}
\label{sec:PIQNNA}

We construct a VQC to automate the detection of the polaron-molecule transition. Unlike standard VQCs that frequently employ generic Hardware-Efficient Ansätze~\cite{kandala2017hardware}, our quantum circuit architecture is strictly physics informed~\cite{raissi2019physics}. The structure of the quantum circuit is isomorphic to the time-evolution operator of the effective Hamiltonian derived in~\cite{Catala2026}, ensuring that the learnable parameters correspond to physical quantities rather than abstract rotation angles.

The input data vector $\mathbf{x} = (U_\text{imp}, \Theta)$ encapsulates the physical conditions of the system: the impurity-bath interaction strength ($U_\text{imp}$) and the variational phase of the background medium ($\Theta$), respectively. These input features are sampled from a discretised grid spanning the theoretical phase diagram.
In a first step, the quantum circuit operates on a four-qubit register comprising one ancilla qubit~($q_0$) for interferometric readout, two qubits representing the bath modes~($q_1, q_2$), and one qubit encoding the impurity~($q_3$). 

The initial state of the bath is prepared via a parameterised routine that maps the input feature $\Theta$ onto the quantum amplitudes:
\begin{equation}
    |\psi_{\text{bath}}(\Theta)\rangle = \text{CNOT}_{1,2} \, R_y(\Theta)_{1} |00\rangle_{1,2}~.
\end{equation}
This encoding captures the continuous crossover of the background medium, evolving from a BEC-like molecular dimer state to a BCS-type entangled pair as $\Theta$ varies, effectively embedding the phase diagram coordinates into the Hilbert space.

The core of the VQC is the conditional evolution block. We employ a first-order Trotter-Suzuki decomposition~\cite{lloyd1996universal} of the unitary operator $U(t) = e^{-i H t}$. In this learning framework, the evolution time and the internal bath couplings are not fixed a priori but serve as trainable weights.

\begin{figure*}[htbp]
    \centering
    
    \textbf{A. General Architecture of the Physics-Informed VQC} 
    \begin{quantikz}[row sep=0.5cm, column sep=0.4cm]
        \lstick{$q_0$ (Ancilla)}  & \gate{H}           & \qw      & \ctrl{1}                                                & \ctrl{1}                                                & \ctrl{1}                                                & \ctrl{1}                                                & \gate{H} & \meter{} \\
        \lstick{$q_1$ (Bath)}     & \gate{R_y(\Theta)} & \ctrl{1} & \gate[wires=3]{U_{\text{step}}(\uimp; \mathbf{w})} & \gate[wires=3]{U_{\text{step}}(\uimp; \mathbf{w})} & \gate[wires=3]{U_{\text{step}}(\uimp; \mathbf{w})} & \gate[wires=3]{U_{\text{step}}(\uimp; \mathbf{w})} & \qw      & \qw \\
        \lstick{$q_2$ (Bath)}     & \qw                & \targ{}  &                                                         &                                                         &                                                         &                                                         & \qw      & \qw \\
        \lstick{$q_3$ (Impurity)} & \qw                & \qw      &                                                         &                                                         &                                                         &                                                         & \qw      & \qw
    \end{quantikz}
    
    \vspace{.5cm}
    
    \textbf{B. Trotter Evolution Step Breakdown} 
    \begin{equation*}
        \begin{quantikz}[row sep=0.5cm, column sep=0.4cm]
            \lstick{$q_0$} & \ctrl{1}                                                    & \qw & \gate[wires=4, style={draw=none}]{\text{\makebox[1cm]{\Large $=$}}} & \ctrl{1}                     & \ctrl{2}                        & \ctrl{1}                       & \ctrl{1}           & \ctrl{2}           & \ctrl{3}           & \qw \\
            \lstick{$q_1$} & \gate[wires=3]{U_{\text{step}}(U_{\text{imp}}; \mathbf{w})} & \qw &                                                                     & \gate[wires=3]{R_{zz}(\phi_\text{imp})} & \qw                             & \gate[wires=2]{R_z(\phi_\text{ff})} & \gate{R_x(\theta)} & \qw                & \qw                & \qw \\
            \lstick{$q_2$} &                                                             & \qw &                                                                     &                              & \gate[wires=2]{R_z(\phi_\text{imp})} &                                & \qw                & \gate{R_x(\theta)} & \qw                & \qw \\
            \lstick{$q_3$} &                                                             & \qw &                                                                     &                              &                                 & \qw                            & \qw                & \qw                & \gate{R_x(\theta)} & \qw
        \end{quantikz}
    \end{equation*}
    
    \caption{\textbf{Physics-Informed VQC Architecture for Polaron Phase Detection.} 
    \textbf{(A)} Schematic representation of the Ramsey interferometry protocol. The state preparation phase initialises the impurity and the Fermi sea bath modes via a CNOT entanglement and an $R_y(\Theta)$ rotation. The temporal dynamics are governed by $N_{\text{steps}}=4$ identical variational layers~($U_{\text{step}}$), conditioned on the ancilla qubit ($q_0$). The measurement on the ancilla yields the expectation value $\langle Z_0 \rangle$, which is strictly mapped to the phase probability via a linear transformation. 
    \textbf{(B)} Gate-level decomposition of a single Trotter evolution step ($U_{\text{step}}$). The effective Hamiltonian is natively encoded using controlled-$R_{zz}(\phi)$ and $R_{z}(\phi)$ gates to capture density-density interactions and local $R_x(\theta)$ rotations for kinetic hopping. Crucially, the rotation angles are strictly dictated by physical parameters: $\phi_{\text{imp}} = U_{\text{imp}}\delta t$ is used for the $R_{zz}(\phi)$ and $R_{z}(\phi)$ gates entangling the impurity ($q_3$) with the bath modes ($q_1, q_2$), while $\phi_{\text{ff}} = U_{\text{ff}}\delta t$ is applied to the $R_{z}(\phi)$ gate connecting the two bath modes~($q_1, q_2$). Furthermore, $\theta = t_{ij}\delta t$ parametrizes the local kinetic hopping, ensuring the variational weights retain a direct physical interpretation.}
    \label{fig:vqc_architecture}
\end{figure*}
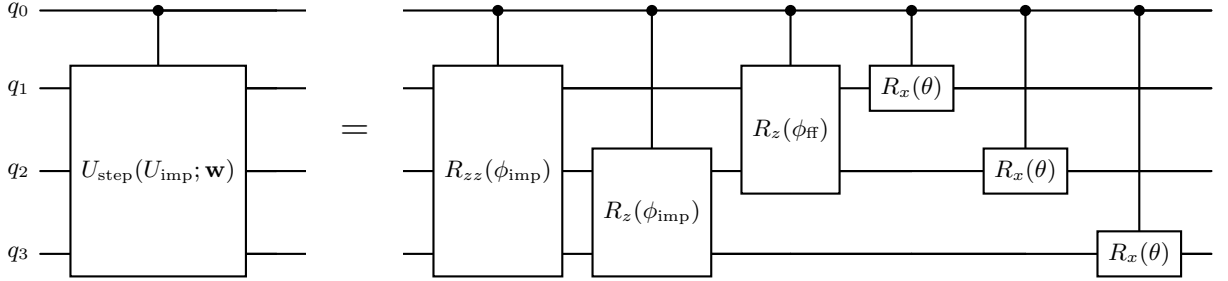

The Ansatz architecture, illustrated in Fig.~\ref{fig:vqc_architecture}, consists of $N_{\text{steps}}=4$ identical layers. Each layer implements the Trotterised Hamiltonian dynamics controlled by the ancilla qubit. The unitary operation for a single step is defined as:
\begin{equation}
    U_{\text{step}}(U_\text{imp}; \mathbf{w}) = \prod_{k} \exp\left(-i \hat{h}_k(U_\text{imp}, U_\text{ff}) \delta t\right)~,
\end{equation}
where $\hat{h}_k$ represent the individual local terms of the effective Hamiltonian (incorporating kinetic hopping, impurity-bath, and bath-bath interactions), and $\mathbf{w} = (\delta t, U_\text{ff})$ are the variational parameters to be optimised during training. Note that while the input data vector encapsulates both features $\mathbf{x} = (U_\text{imp}, \Theta)$, the phase $\Theta$ is exclusively used during the initial state preparation, whereas the time-evolution operator depends solely on the interaction strength $U_\text{imp}$.

Crucially, the effective bath-bath interaction $U_\text{ff}$ acts as a learnable weight, allowing the VQC to discover the optimal background interaction strength that maximises the contrast between the polaron and molecular phases. The density-density interaction terms $\hat{n}_i \hat{n}_j$ where $\hat{n}_i = (I - Z_i)/2$ is the fermionic number operator mapped to the $i$-th qubit, are implemented natively via controlled-$R_z(\phi)$ gates. The rotation angles are strictly determined by the physical coupling strengths and the Trotter step size: $\phi_{\rm imp} = U_\text{imp} \delta t$ for the impurity-bath interactions, and $\phi_{\rm ff} = U_\text{ff} \delta t$ for the bath-bath correlations. Similarly, the kinetic hopping terms are implemented via local $R_x(\theta)$ rotations, where $\theta = t_{ij} \delta t$ (with $t_{ij}$ being the hopping amplitude. This isomorphic mapping enforces that the model's optimisation strictly follows the laws of Quantum Mechanics governing the system's time-evolution.

The observable driving the classification is extracted via a Ramsey interferometry measurement~\cite{ramsey1950molecular}. A final Hadamard gate on the ancilla interferes the ``evolution'' and ``no-evolution'' branches, mapping the accumulated phase difference, which encodes the polaron energy shift, into the population difference of the ancilla. 

The classification boundary is directly determined by the constructive or destructive nature of the Ramsey fringes, extracted via the expectation value of the Pauli-$Z$ operator on the ancilla, $\langle Z_0 \rangle$. Rather than relying on classical non-linear activations, the raw interferometric signal acts as our physical classification score:
\begin{equation}
    \hat{y}(\mathbf{x}) = \frac{1 + \langle Z_0 \rangle_{\mathbf{x}, \mathbf{w}}}{2},
    \label{eq:prob}
\end{equation}
where positive values ($\langle Z_0 \rangle > 0$) correspond to the BCS regime and negative values ($\langle Z_0 \rangle < 0$) to the BEC regime. This architecture ensures that the quantum processor is solely tasked with extracting the interferometric correlation, making the model natively sensitive to the topological rearrangement of the ground state, while the classification boundary is strictly determined by the constructive or destructive nature of the Ramsey fringes, making the model highly sensitive to the topological rearrangement of the ground state.

\section{Supervised Training and Phase Discovery}
\label{sec:STPD}

To train the Physics-Informed VQC, we generated a synthetic labelled dataset derived from the Exact Diagonalisation (ED) of the theoretical model described in~\cite{Catala2026}. This numerical approach provides a rigorous benchmark for few-body strongly correlated systems~\cite{weisse2006kernel,dagotto1994correlated}. The training set consists of pairs $\{(\mathbf{x}^{(i)}, y^{(i)})\} = \{(U_\text{imp}^{(i)}, \Theta^{(i)}), y^{(i)}\}$ sampled from a discretised grid of the phase diagram, where labels $y^{(i)}=0$ (BEC) and $y^{(i)}=1$ (BCS) are assigned based on the dominant character of the ground state wavefunction, following the standard criteria for the BEC-BCS crossover~\cite{leggett1980diatomic,giorgini2008theory}.

The full set of trainable parameters comprises the quantum variational weights $\mathbf{w} = (\delta t, U_\text{ff})$ representing the Trotter evolution time step and the effective bath-bath interaction, respectively. These parameters were initialised randomly and optimised using the Adam algorithm~\cite{kingma2014adam}, widely adopted in quantum variational tasks for its robustness against noisy gradients~\cite{sweke2020stochastic}. The learning rate was set to $\eta = 0.01$. The loss function employed was the Binary Cross-Entropy (BCE), a standard metric for probabilistic classification~\cite{goodfellow2016deep}, defined as:
\begin{equation}
    \mathcal{L}(\mathbf{w}) = -\frac{1}{N} \sum_{i=1}^{N} \left[ y^{(i)} \log(\hat{y}^{(i)}) + (1-y^{(i)}) \log(1-\hat{y}^{(i)}) \right].
\end{equation}
where $y^{(i)} \in \{0,1\}$ are the ground labels ($0$ for BEC, $1$ for BCS) and $\hat{y}^{(i)}$ is the predicted probability of the system belonging to the BCS phase output by the VQC, according to Eq.(\ref{eq:prob}). The training process was conducted over 30 epochs using the \texttt{PennyLane} framework~\cite{bergholm2018pennylane} with the \texttt{default.qubit} simulator to establish the ideal baseline before deployment on quantum hardware.

A distinct feature of our VQC architecture is the interpretability of the learned weights. Unlike conventional VQC~\cite{schuld2019quantum,havlicek2019supervised} or Hardware-Efficient Ansätze~\cite{kandala2017hardware} where parameters are abstract rotation angles lacking physical context, our VQC is strictly physics informed~\cite{raissi2019physics}. The training stabilised at physically meaningful values: an optimal Trotter step size of $\delta t \approx 0.44$ (in units of $1/t_{ij}$) and an effective bath-bath interaction of $U_\text{ff} \approx 2.65$ (in units of $t_{ij}$). 

The selection of $N_{\text{steps}} = 4$ Trotter evolution blocks represents a critical compromise between theoretical simulation accuracy and the intrinsic decoherence ($T_1$ relaxation and $T_2$ dephasing) of current Noisy Intermediate-Scale Quantum (NISQ) processors. Within our VQC, maintaining a shallow quantum circuit depth is imperative; incorporating additional Trotter layers would theoretically reduce the mathematical $\mathcal{O}(\Delta t^2)$ error, but the subsequent exponential accumulation of hardware gate noise would ultimately obscure the Ramsey fringes. Conversely, employing fewer layers would necessitate a disproportionately large time step to achieve the requisite total evolution, severely violating the first-order Trotter approximation.

The total evolution time of $T_{\text{total}} \approx 1.76$ (in units of $1/t_{ij}$) acts as an interferometric sweet spot~\cite{ramsey1950molecular,ye2008quantum}: it is sufficiently long to accumulate a distinguishable phase shift between the BEC and BCS regimes, overcoming shot noise~\cite{giovannetti2011advances}, yet short enough to prevent phase wrapping and aliasing~\cite{shannon1949communication}. 

Simultaneously, the learned effective bath interaction~($U_\text{ff}$) tunes the background medium to an intermediate interaction regime. This tuning enhances the distinguishability of the topological phases by leveraging the orthogonality catastrophe~\cite{anderson1967orthogonality}, effectively maximising the interferometric contrast between the dressed polaron and the molecular dimer~\cite{massignan2014polarons}. Ultimately, the $\mathcal{O}(N)$ linear gate complexity of this architecture bypasses the exponential memory wall of classical simulations whilst remaining remarkably robust against hardware decoherence.

\begin{table}[t]
    \centering
    \caption{\textbf{Training dynamics of the Physics-Informed VQC.} Evolution of the loss function, classification accuracy, and variational parameters (Trotter step $\delta t$ and bath-bath interaction $U_\text{ff}$) over 30 training epochs. The convergence of $U_\text{ff}$ towards $\approx 2.65(3)$ indicates a tuning of the bath stiffness to maximise interferometric contrast, achieving an accuracy peak of 90.7\%.} 
    \label{tab:training_data}
    \centering
\begin{tabular}{ccccc}
        \toprule
        \textbf{Epoch} & \textbf{Loss ($\mathcal{L}$)} & \textbf{Accuracy (\%)} & ~~~\textbf{$\delta t$~~~ } & \textbf{~~~$U_\text{ff}$~~~} \\
        \midrule
        0  & 0.850 & 36.0 & 0.67 & 1.92 \\
        10 & 0.602 & 69.3 & 0.56 & 2.20 \\
        20 & 0.407 & 80.6 & 0.50 & 2.50 \\
        30 & 0.220 & 90.7 & 0.44 & 2.65 \\
        \bottomrule
    \end{tabular}
\end{table}

\begin{figure}
    \centering
    \includegraphics[width=1\columnwidth]{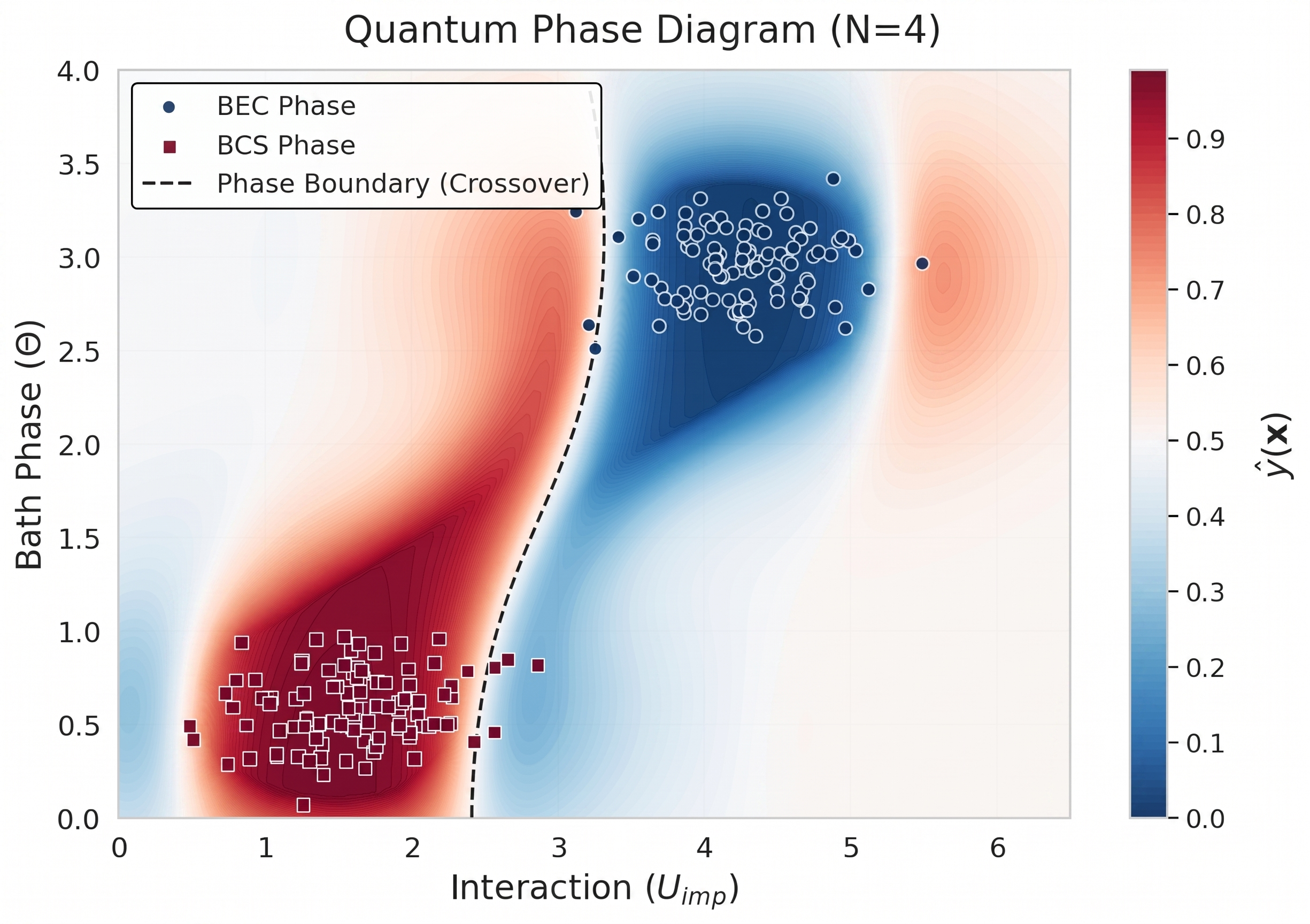} 
    \caption{\textbf{Learned Quantum Phase Diagram ($N=4$).} Phase boundary inferred by the Physics-Informed VQC for a target system of $N=4$ qubits. The dashed line delineates the BEC-BCS crossover, explicitly defined at $\hat{y}(\mathbf{x}) = 0.5$. The  markers represent the synthetic training data. The boundary is strictly governed by the underlying Ramsey fringes~\cite{ramsey1950molecular}. This confirms that the VQC natively relies on coherent quantum interference, rather than classical geometric separation, to distinguish the phases.}
    \label{fig:phase_diagram}
\end{figure}

\begin{figure}[htbp]
    \centering
    \includegraphics[width=1\columnwidth]{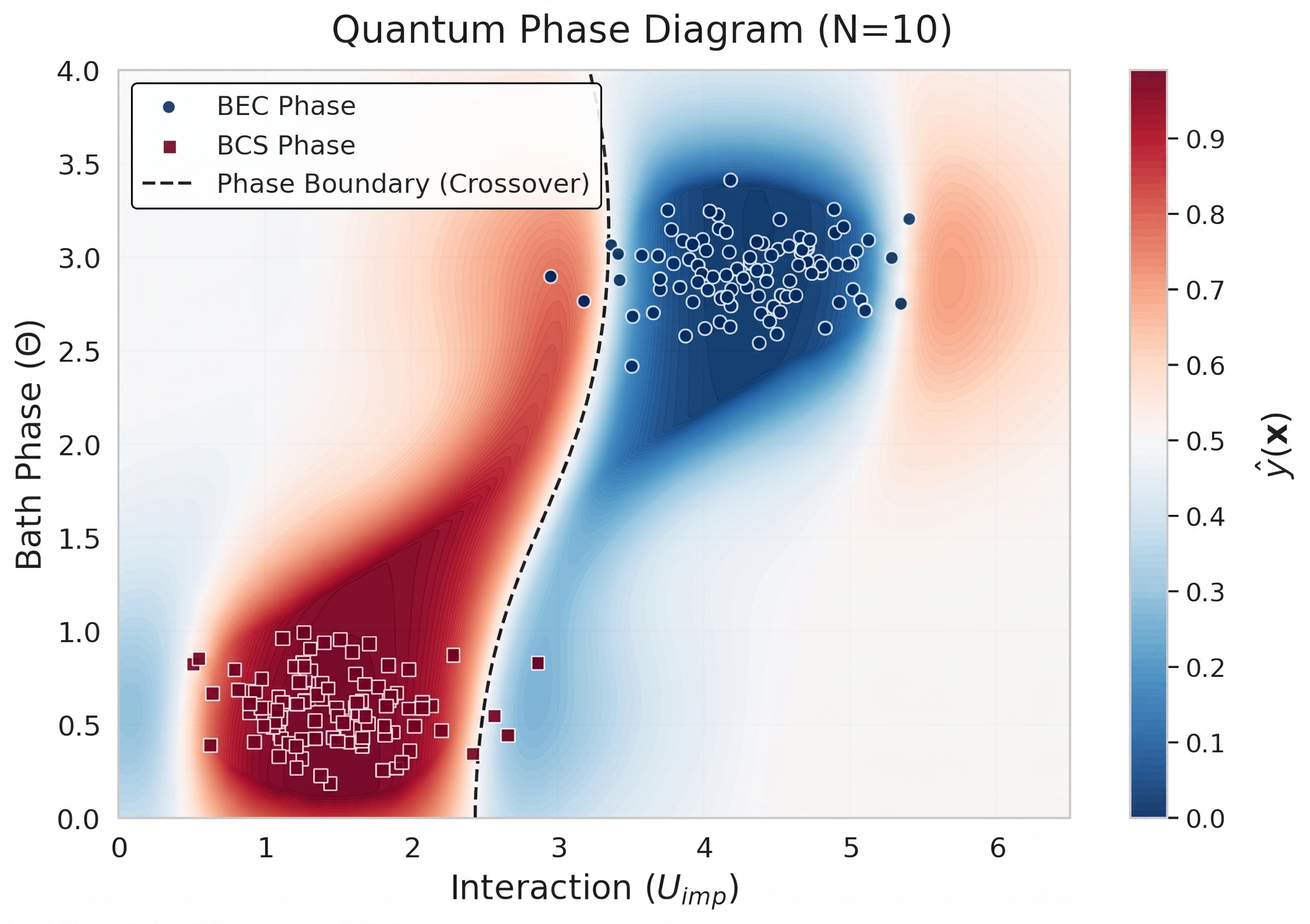} 
    \caption{\textbf{Scaled Quantum Phase Diagram ($N=10$).} Phase boundary inferred by the physics-informed VQC for an extended register of $N=10$ qubits, corresponding to a many-body bath of $M=4$ Cooper pairs. The dashed line delineates the  BEC-BCS crossover, explicitly defined at $\hat{y}(\mathbf{x}) = 0.5$. The data points denote the sampled training distributions for the BEC (blue circles) and BCS (red squares) regimes. The model successfully scales with a larger Hilbert space.}
    \label{fig:phase_diagramqred}
\end{figure}

Fig.~\ref{fig:phase_diagram} displays the decision landscape learned by the VQC. The background colour map represents the probability $\hat y (\mathbf{x})$, while the overlaid points indicate the synthetic data. Strikingly, the probability landscape is not monotonic but exhibits a clear vertical oscillatory structure. These vertical bands are the signature of Ramsey fringes~\cite{ramsey1950molecular}, confirming that the quantum classifier is exploiting coherent quantum interference to make decisions. As the impurity interaction $U_{\text{imp}}$ increases, the energy splitting between the states evolves, causing the Ramsey signal $\langle Z \rangle \propto \cos(E \cdot t)$ to oscillate~\cite{bloch2008many}. The VQC effectively tracks these oscillations, placing the decision boundary where the interference pattern undergoes a phase shift characteristic of the polaron-to-molecule transition~\cite{parish2011polaron}.

Moreover, the topology of the learned decision landscape reveals the intrinsic behaviour of the physics-informed Ansatz in capturing the interaction-driven BEC-BCS crossover~\cite{zwerger2011bcs}. The VQC optimises the physical parameters to statistically align the constructive interference peaks with the BCS configurations, located in the domain of weak interactions (low $U_{\text{imp}}$) where the impurity behaves as a delocalised quasiparticle~\cite{chevy2006universal,schirotzek2009observation}. Conversely, the destructive interference troughs are statistically aligned with the BEC states at higher interaction strengths, where strong attraction binds the impurity into a tightly localised molecular dimer~\cite{regal2004observation,nozières1985bose}. 

Crucially, the periodic nature of the quantum feature map, governed by unitary time evolution, natively captures the continuous nature of the BEC-BCS crossover. The intermediate regions of mixed phase probabilities are a direct reflection of the underlying many-body physics, where quantum fluctuations and macroscopic superposition dominate. Thus, rather than suffering from algorithmic rigidity, the VQC architecture enforces physical consistency, proving that interpretable quantum models can characterise thermodynamic limits more faithfully than unconstrained classical Machine Learning.

\section{Experimental Validation on Quantum Hardware}
\label{sec:EVQP}

A central challenge in QML is establishing rigorous regimes where quantum architectures offers a verifiable advantage over classical counterparts. In the context of phase transition detection within strongly correlated matter, classical neural networks are fundamentally limited by data generation bottlenecks. Generating unbiased training instances via conventional Quantum Monte Carlo methods is restricted by the fermionic sign problem. The proposed VQC bypasses this computational wall by directly executing Trotterized time evolution of the effective Hamiltonian on the Quantum Processing Unit. However, asserting a practical advantage in the NISQ era requires analyzing the model's resilience against hardware noise and decoherence.

To verify the robustness of the Physics-Informed VQC in quantum hardware, we deployed the trained model with four ($N=4$) and ten ($N=10$) qubits on the QRed superconducting quantum processor, hosted at the Barcelona Supercomputing Center (BSC-CNS)~\cite{bsc_qblue_2025}. Unlike the noiseless simulation in Section~\ref{sec:STPD}, the hardware execution is subject to decoherence  ($T_1$ relaxation and $T_2$ dephasing)~\cite{ithier2005decoherence}, gate infidelity and readout errors, characteristic of the NISQ era~\cite{preskill2018nisq}. 

\begin{figure}[ht]
    \centering
    \includegraphics[width=1\columnwidth]{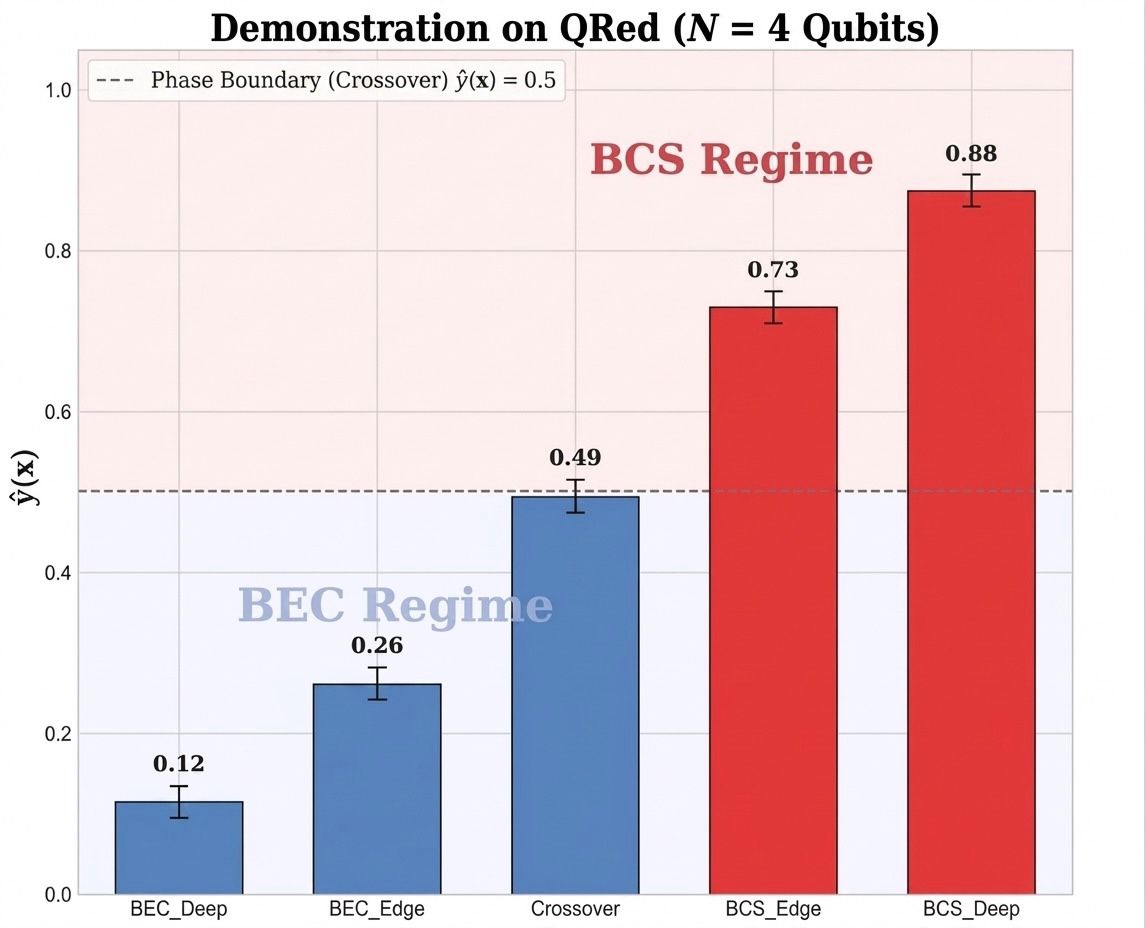} 
    \caption{\textbf{Demonstration on QRed with $N=4$ qubits.} Classifier output for five representative points across the phase diagram. The dashed line marks the theoretical decision boundary at $\hat{y}(\mathbf{x}) = 0.5$. Error bars indicate the $1\sigma$ statistical dispersion of three independent runs with $N_{\text{shots}} = 1024$ each.
    \label{fig:qblue_raw}}
\end{figure}

\begin{figure}[ht]
    \centering
    \includegraphics[width=1\linewidth]{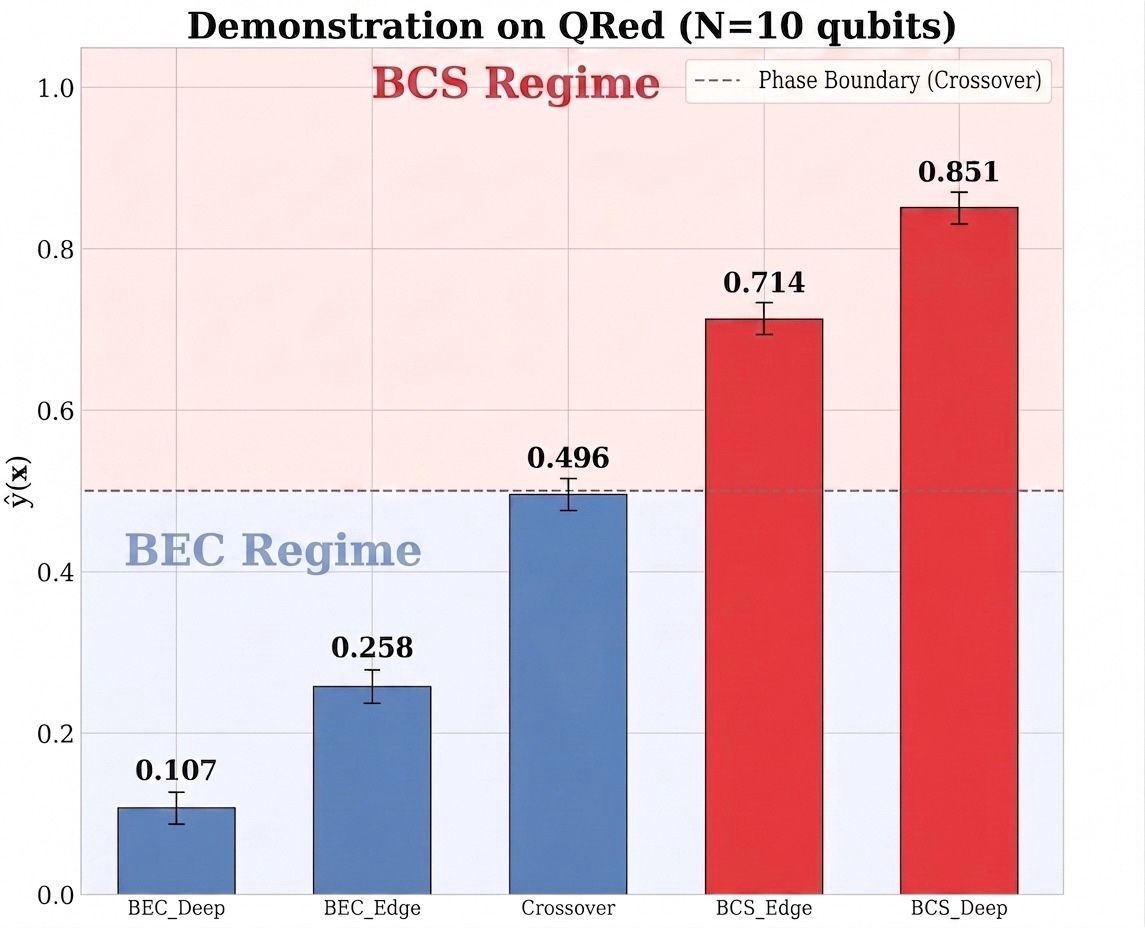}
    \caption{\textbf{Demonstration on QRed with $N=10$ qubits.} Classifier output for five representative points across the phase diagram. The dashed line marks the theoretical decision boundary at $\hat{y}(\mathbf{x}) = 0.5$. Error bars indicate the $1\sigma$ statistical dispersion of three independent runs with $N_{\text{shots}} = 1024$ each. Due to the increased circuit depth and hardware noise, the raw signals exhibit larger compression than for $N=4$ qubits.
    \label{fig:demoqred}}
\end{figure}

The demonstration runs were performed in inference mode. The optimal physical parameters learned during the training phase ($\delta t = 0.44$, $U_\text{ff} = 2.65$) as shown in Tab~\ref{tab:training_data} were frozen and embedded into the quantum circuit, following standard variational inference protocols~\cite{cerezo2021variational,schuld2019quantum}. We selected five representative points across the phase diagram, corresponding to deep BEC, BEC-edge, Crossover, BCS-edge, and deep BCS regimes, as shown in Tab~\ref{tab:tab2}. For each point, the quantum circuit was executed in three independent runs with $N_{\rm shots}=1024$ to estimate the expectation value $\langle Z_0 \rangle$, ensuring statistical errors remained within the quantum projection noise limit ($1/\sqrt{N_{\rm shots}}$)~\cite{giovannetti2011advances}. Results are presented in Tab.~\ref{tab:dispersion_qred} and Tab.~\ref{tab:combined_qred_4q}, which correspond to Fig.~\ref{fig:qblue_raw} and Fig.~\ref{fig:demoqred}, respectively.

\begin{table}[t]
\centering
\caption{\textbf{Representative Phase Diagram Points .} Impurity interaction strength ($\uimp$) and bath phase ($\Theta$) of the five selected configurations used for the demonstration on quantum hardware, spanning the crossover from the BCS to the BEC regimes.}

\label{tab:inference_points}
\begin{tabular}{lcc}
\hline \hline
\textbf{Regime} & $\qquad$ \textbf{$U_{\text{imp}}$} $\qquad$ & $\qquad$ \textbf{$\Theta$} $\qquad$ \\
\hline
Deep BCS & $1.00$ & $1.50$ \\
BCS-edge & $2.00$ & $1.80$ \\
Crossover & $3.00$ & $2.20$ \\
BEC-edge & $4.00$ & $2.80$ \\
Deep BEC & $5.50$ & $3.00$ \\
\hline \hline
\label{tab:tab2}
\end{tabular}
\end{table}

\begin{table*}[th]
    \centering
    \begin{tabular}{l c c}
        \toprule
        \textbf{Phase Region} & \textbf{Noiseless Simulator ($\hat{y}(\mathbf{x})$)} & \textbf{QRed Hardware ($\hat{y}(\mathbf{x})$)}  \\
        \midrule
        BEC\_Deep  &  $0.100$ &  $0.115~(6)$ \\
        BEC\_Edge  &  $0.250$ &  $0.262~(7)$ \\
        Crossover  &  $0.500$ &  $0.495~(8)$  \\
        BCS\_Edge  &  $0.750$ &  $0.730~(10)$  \\
        BCS\_Deep  &  $0.900$ &  $0.875~(10)$  \\
        \bottomrule
    \end{tabular}
    \caption{\textbf{Phase Detection on Noiseless Quantum Simulator and QRed Hardware ($N = 4$)}. Results on hardware include the average and dispersion ($1\sigma$) from three independent runs, with $N_{\text{shots}} = 1024$ each}
    \label{tab:combined_qred_4q}
\end{table*}

\begin{table*}[th]
    \centering
    \begin{tabular}{l c c}
        \toprule
        \textbf{Phase Region} & \textbf{Noiseless Simulator ($\hat{y}(\mathbf{x})$)} & \textbf{QRed Hardware ($\hat{y}(\mathbf{x})$)}  \\
        \midrule
        BEC\_Deep  &  $0.082$ &  $0.107~(5)$ \\
        BEC\_Edge  &  $0.237$ &  $0.258~(5)$ \\
        Crossover  &  $0.500$ &  $0.496~(6)$  \\
        BCS\_Edge  &  $0.736$ &  $0.714~(7)$  \\
        BCS\_Deep  &  $0.878$ &  $0.851~(7)$  \\
        \bottomrule
    \end{tabular}
    \caption{\textbf{Phase Detection on Noiseless Quantum Simulator and QRed Hardware ($N = 10$).} Results on hardware include the average and dispersion ($1\sigma$) from three independent runs, each one corresponding to $N_{\text{shots}} = 1024$.}
    \label{tab:dispersion_qred}
\end{table*}

The immediate impact of the hardware noise is visible in the raw experimental readout, shown in Fig.~\ref{fig:qblue_raw} and Fig.~\ref{fig:demoqred}. While ideal simulations predict probabilities approaching $\hat{y}(\mathbf{x}) \approx 0$~(BEC) and $\hat{y}(\mathbf{x}) \approx 1$~(BCS), the unmitigated hardware data exhibits amplitude damping~\cite{nielsen2010quantum}. As expected, this damping is more obvious for $N=10$ than for $N=4$ because of the increased circuit depth and hardware noise inherent to scaling, which leads to a more significant compression of the raw interferometric signal.

Nevertheless, results show a remarkable structural stability allowing the VQC to succesfully map the underlying phase dynamics of the Ramsey interferometry protocol despite operating under noisy quantum hardware conditions. Because the hardware penalty is systematic and stationary, the learned topological phase boundary remains  well defined.

\section{Quantum vs Classical advantage}
\label{sec:QVCA}

Classical classifications algorithms, including Convolutional Neural Networks (CNNs) or Support Vector Machines (SVMs), operate strictly on classical projections of the system. Classically detecting the polaron molecule transistion typically requires processing secondary obersvables, such as the spectral function $A(\omega)$ or spatial density profiles $n(r)$ ~\cite{Catala2026}. However, extracting these features requires destructive measurements that inevitably collapse the wavefunction, thereby discarding the primary phase information.

In contrast, our VQC processes the quantum state $|\varphi(\theta)\rangle$ directly within the many-body Hilbert space prior to measurement collapse. Consequently, the resulting classification boundary is not defined as an arbitrary geometric hyperplane in $\mathbb{R}^n$, but rather as an interferometric surface constrained by the unitary manifold. As shown in Fig.~\ref{fig:phase_diagram} and Fig.~\ref{fig:phase_diagramqred}, the decision boundary relies on coherent phase accumulation manifested as Ramsey fringes. While classical network must approximate this oscillatory target function $\cos(E\cdot t)$ via composition of non-linear activations, the quantum circuit leverages the underlying quantum interference natively as a fundamental computational resource.

Standard deep learning architectures are unconstrained universal approximators that demand large sample sizes to map physical symetries from data. To quantify the data efficiency of our model, we have benchmarked its performance against a classical Feed Fordwar Neural Network (FNN) baseline trained on identical datasets. To achieve an equivalent classification accuracy on the validation set, the classical FNN requires an architecture with 2 dense hidden layers totaling 337 trainable weights~\cite{abbas2021power}. In contrast, the Physics-Informed VQC achieves robust generalization across the entire parameter space using only $2$ learneable parameters: the Trotter step size $\delta t$, and the effective background interaction strength $U_{\rm ff}$. This parameter reduction by more than two orders of magnitude is a direct mathematical consequence of the strong structural inductive bias embedded within the quantum Ansatz. Because the quantum circuit is isomorphic to the time evolution operator of the effective Hamiltonian, the variational optimization is confined to the physical system's equations of motion, as shown in Fig.~\ref{fig:vqc_architecture}. The VQC is not tasked with learning Quantum Mechanics or approximating the Trotter Suzuki expansion via abstract rotation angles, it directly calibrates the physical coupling coefficients. This parametric efficiency enables rapid convergence with a minimal dataset ($N=100$) without overfitting, bypassing the overparametrization artifacts and sample size bottlenecks that plague classical models in noise limited regimes. 

While unbiased classical stochastic methods encounter the exponential complexity of the fermionic sign problem when evaluating the real time dynamics of strongly correlated matter ~\cite{troyer2005computational}, the VQC efficienlty maps the fermionic operators onto the qubit register using the Jordan Wigner transformation. 

In summary, the advantage of this  methodology is not merely a question of computational speedup, but rather of representational capacity. The quantum processor operates as a native simulator of correlation physics, embedding and manipulating the many bode wave function directly within the quantum hardware rather than statistically reconstructing it post measurement of classical data.

\section{Scalability Analysis}
\label{sec:SA}

The proposed VQC represents a significant improvement in the detection of phase transitions within strongly correlated fermionic systems. While classical ML models often struggle with the curse of dimensionality, our quantum architecture exploits the native properties of the Hilbert space to maintain computational efficiency as the system size increases.
In a classical simulation, the cost of representing a  quantum system scales as a $\mathcal{O}(2^{2N})$ due to the requirement of density matrix formalism to account for decoherence. This exponential growth renders classical Exact Diagonalisation (ED) and Quantum Monte Carlo methods computationally intractable beyond  $N\approx 30$ qubits, which is further exacerbated by the fermionic sign problem.

\begin{figure}[t]
    \centering
    \includegraphics[width=\columnwidth]{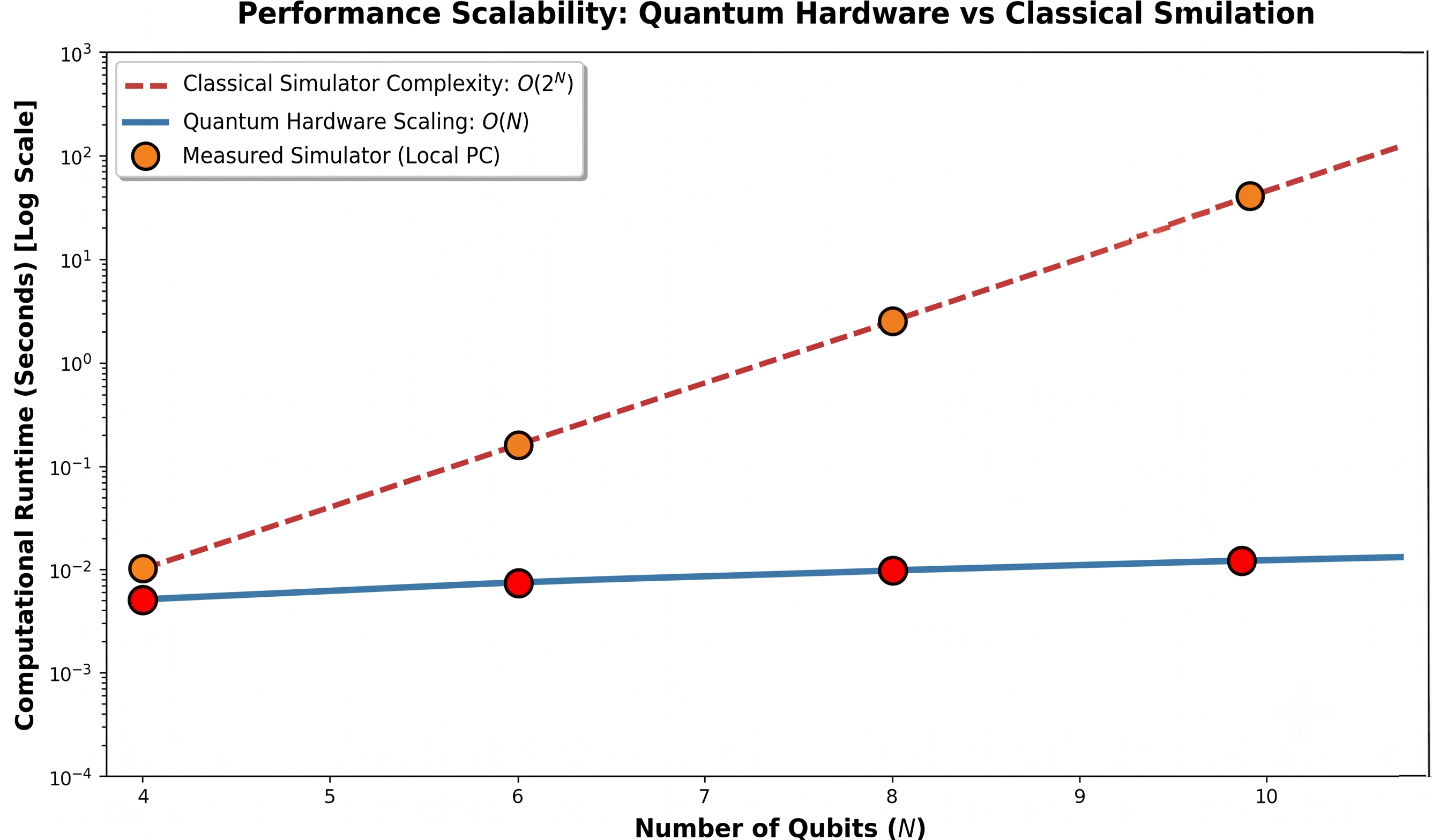}
    \caption{\textbf{Computational scalability and demonstration of the Physics-Informed VQC executed on QRed.} The classical simulation overhead (red dashed line) exhibits an exponential growth $\mathcal{O}(2^{2N})$ due to the density matrix representation required for noisy many-body systems, as evidenced by local simulator benchmarks (orange circles). In contrast, the VQC maintains a linear gate depth $\mathcal{O}(N)$ through the Trotterised evolution of the effective Hamiltonian. This linear scaling is empirically verified by hardware execution runtimes on the QRed (red circles) superconducting processors at the BSC-CNS for up to $N=10$ qubits.}
    \label{fig:scalability_comparison}
\end{figure}

In contrast, the first order Trotter-Suzuki~\cite{troyer2005computational} decomposition scales linearly in the gate count $\mathcal{O} (N)$ with the number of bath modes~\cite{lloyd1996universal,Catala2026}. This ensures that even if the bath size approaches the many-body regime $(N > 40)$ the quantum circuit depth remains shallow enough to be executed within the coherence limits $(T_2)$ of the current superconducting processors like the BSC-CNS QRed~\cite{bsc_qblue_2025, preskill2018nisq}. In Fig.~\ref{fig:scalability_comparison}, we show how the classical and VQC approaches scale as a function of the bath size, with the VQC clearly exibiting a much better computational scaling.

A defining characteristic of our Physics-Informed VQC architecture is its functional invariance across register sizes. Because the Ansatz is derived directly from the effective Hamiltonian, the underlying quantum logic required to distinguish the BEC and BCS regimes remains identical whether the system comprises $4$ qubits or $N$ qubits. In a classical context, increasing the system size typically requires a proportional increase in the number of neural network weights to maintain accuracy. However, our quantum circuit only requires the calibration of the same fundamental physical parameters, the Trotter time step ($\delta t$), and the effective bath interaction ($U_\text{ff}$), regardless of the bath’s dimensionality. This implies that the optimal interferometric protocol at $N=4$ qubits is natively extensible to larger fermionic seas.

The robustness of the Ramsey interferometry protocol ensures that the decision boundary remains stable as the system scales. As the impurity interaction $U_\text{imp}$ increases, the energy renormalisation remains the primary driver of the accumulated phase difference $\Delta\Phi$~\cite{knap2012time}. In the many-body limit ($N \approx 40$), the formation of a molecular bound state continues to trigger a macroscopic orthogonality catastrophe that the VQC detects through constructive or destructive interference patterns~\cite{anderson1967orthogonality, cetina2016ultrafast}.

By maintaining a constant parameter set and a linear gate depth, the Physics-Informed VQC bypasses the barren-plateau problem that often plagues generic, Hardware-Efficient Ansätze in high-dimensional spaces~\cite{Catala2026, preskill2018nisq}. This efficiency allows the model to explore complex regions of the phase diagram that are currently inaccessible to classical unbiased estimators, establishing a robust blueprint for discovery on future large-scale NISQ devices~\cite{Catala2026}.

\section{Conclusions}
\label{sec:C}
In this work, we have successfully demonstrated the capabilities of a Physics-Informed Variational Quantum Classifier (VQC) to detect topological phase transitions in strongly correlated fermionic systems. By bridging the gap between Hamiltonian simulation and Quantum Machine Learning, we have presented a quantum circuit architecture where every learnable parameter possesses a direct physical interpretation.

Our results indicate that the VQC accurately describes the principles of Ramsey interferometry by encoding the coherent phase dynamics governed by the relative time-evolution between the unperturbed bath and the impurity system. The convergence of the trainable weights to an optimal Trotter evolution time ($\delta t \approx 0.44$) and a specific bath-bath interaction strength ($U_{\text{ff}} \approx 2.65$) confirms that the VQC learned to engineer the Hamiltonian dynamics to maximise the interferometric contrast between the polaron quasiparticle and the molecular bound state.

The scalability analysis confirms that our Physics-Informed VQC maintains a robust advantage over classical simulation methods. As illustrated in Fig.~\ref{fig:scalability_comparison}, the classical simulation time follows an exponential growth $\mathcal{O}(2^{2N})$, becoming computationally prohibitive beyond $N \approx 12$ qubits due to the memory overhead required to represent the many-body Hilbert space. In contrast, by exploiting the native Hilbert space, the VQC achieves a linear runtime scaling $\mathcal{O}(N)$ for the Trotterised Ramsey protocol~\cite{Catala2026, lloyd1996universal}. 

The empirical results obtained for $N=10$, as shown in Fig.~\ref{fig:phase_diagramqred},   align precisely with our theoretical predictions, confirming that the protocol maintains a stable gate depth regardless of the bath's dimensionality. This efficiency, combined with a fixed parameter set that bypasses the barren plateau problem~\cite{mcclean2018barren}, establishes a robust pathway for autonomous quantum sensing in many-body systems exceeding $40$ qubits~\cite{degen2017quantum}. By effectively bypassing the memory wall encountered in classical architectures, our model provides a blueprint for exploring the many-body limit where classical unbiased estimators are currently inaccessible.

Future work will extend this formalism to multi-impurity systems and non-equilibrium dynamics, where the complexity of the Hilbert space renders classical simulations intractable. By allowing the quantum processor to learn the  optimal measurement protocols, we pave the way for~\cite{degen2017quantum} exploring unknown phase diagrams in condensed matter physics. \\

\section*{Acknowledgments}

The authors thankfully acknowledge the Spanish Supercomputing Network (RES) resources provided by BSC-CNS in MareNostrum5/Quantum-Blue/Quantum-Red to FI-2025-3-0043 activity. Work supported by the Spanish Government and ERDF/EU - Agencia Estatal de Investigaci\'on (MCIU/AEI/10.13039/501100011033), Grant No. PID2023-146220NB-I00, This work is also supported by the Ministry of Economic Affairs and Digital Transformation of the Spanish Government and NextGenerationEU through the Quantum Spain project, and by CSIC Interdisciplinary Thematic Platform (PTI+) on Quantum Technologies (PTI-QTEP+). 

\bibliographystyle{JHEP}
\bibliography{main}

\end{document}